\documentclass{iaus260}
\usepackage{graphicx}
\pubyear{2009}
\volume{260}  
\pagerange{10--17}
\jname{The R\^ole of Astronomy in Society and Culture}
\editors{D. Valls-Gabaud \& A. Boksenberg, eds.}

\title[Astronomy, Space Science and Geopolitics]                   
{Astronomy, Space Science and Geopolitics}  

\author[Thierry J.-L. Courvoisier]           
{Thierry J.-L. Courvoisier}        

\affiliation{ISDC Data centre for Astrophysics, University of Geneva \\ 
                 16 ch. d'Ecogia, CH-1290 VERSOIX, Switzerland \\ 
                 email: {\tt thierry.courvoisier@unige.ch} \\[\affilskip]
             }

\begin{document}
\maketitle

\begin{abstract}
Astronomy has played a major part in the development of civilisations, not only through conceptual developments, but most importantly through the very practical gains obtained through the observation of Sun, Moon planets and stars. Space sciences, including astronomy,  have also played a major role in the development of modern societies, as engine for most subsequent space technology developments. Present trends tend to decrease the role of science in space development. This trend should be reversed to give modern "societies" their independence  in space related matters that permeate the lives of all inhabitants of the Earth.    
\keywords{Space science, international development}    
\end{abstract}

\firstsection 
\section{Astronomy in our societies}

We, astronomers and people around us now, tend to see in astronomy an activity that is mainly, if not solely devoted to the understanding of the Universe and the objects to be found within. This is an intellectual quest that is particularly fruitful since the beginning of the space age and the opening of the electro-magnetic spectrum from radio waves to gamma rays to the observations of the sky. This is certainly a valid point of view now. The quest for understanding as an activity for itself  has, however, certainly not been the main activity of astronomers over history.

Most of the astronomical workforce has been invested not so much in a cultural endeavour, but rather in a tedious quest for the measurement and keeping of time. Agriculture requires that seeds are planted well in advance of the season of growth. It is, therefore, necessary to know when to plant. This time cannot be estimated on the current weather, but requires advance planning for which the only useful information is based on the positions of Sun, Moon and stars. Since the Moon month and the year do not have a simple relation, the quest for timekeeping during the year is a complex one. Astronomers have, therefore, spent large efforts world wide to solve this problem. 

The same is true for other needs of society, navigation  certainly requires seaworthy  ships, it also requires the capacity to locate oneself on the surface of the Earth in unknown territories and on sea. Again, this is provided by a knowledge of the respective positions of Sun, Moon, planets and stars, together with a good mastering of time keeping. All of this knowledge is based on astronomy.

The daily needs of human society also require some level of time keeping, be it only to be able to meet at given time and place. Here again the local timekeeping in many organised societies has relied on astronomical observations. The results being then relayed to the population by bells, canons and other signals.

Astronomy has thus been a most practical endeavour for most of the human history. The present conference amply demonstrates this not only in the European culture, but also  in other ones. The benefits of astronomy for society cannot be overestimated. Agriculture, navigation and the organisation of societies have depended crucially on astronomical observations for almost the whole of the development of human civilisations. The development of a "Weltbild" should be seen in this context as an, important, side benefit.

The practical importance of astronomy has only rather recently ceased to dominate our activities. The observatory of Geneva has, as one example, been in charge of certifying chronometers for the local manufactures until  the late 1960's. It is also interesting to learn that the observatory of Besan\c{c}on, in France, North of the Jura mountains, was founded   in 1878  not so much to contribute to the great human quest for knowledge about the world than to help the local watch manufacturers who considered themselves at a disadvantage compared to their competitors of the South of the Jura mountains, in Neuch\^atel and Geneva, who had access to astronomical observatories.

It is, therefore, only very recently that Astronomy as become a mainly cultural quest, the practical benefits of which are only minor and indirect, to be found in the knowledge gained in making complex instruments capable of observing in remote and hostile environments.

\section{Space Sciences and The Development of Modern Societies}

\subsection{The Past}

Space sciences have been a prime mover, eventhough not the only one, in the development of space activities. Astronomy, as a major part of space sciences,  has thus had a leading role in the build up of space capacities in the different parts of the world that developed them in the last decades. This role, as well as the more nationalistic connotations of the development, is well illustrated by the following section of the Moon speech given by president Kennedy in Houston in 1962:

{\it
"Those who came before us made certain that this country rode the first waves of the industrial
revolutions, the first waves of modern invention, and the first wave of nuclear power, and this
generation does not intend to founder in the backwash of the coming age of space. We mean to
be a part of it--we mean to lead it. For the eyes of the world now look into space, to the moon and
to the planets beyond, and we have vowed that we shall not see it governed by a hostile flag of
conquest, but by a banner of freedom and peace. We have vowed that we shall not see space
filled with weapons of mass destruction, but with instruments of knowledge and understanding.

Yet the vows of this Nation can only be fulfilled if we in this Nation are first, and, therefore, we
intend to be first. In short, our leadership in science and in industry, our hopes for peace and
security, our obligations to ourselves as well as others, all require us to make this effort, to solve
these mysteries, to solve them for the good of all men, and to become the world's leading spacefaring
nation."
}

In short, let space be devoted to peaceful scientific activities, but let this development be a US development.

In Europe the role of science in the development of space activities has even been more prominent. This is certainly due to the fact that Europe is not, yet?, identified as such in a nationalistic sense. National pride or interests have therefore had less influence than elsewhere around the Earth.

Since the early developments of the 1960's, space activities have become a very important element of the infrastructures of our societies. Their importance is often not quite valued at the right level. Suffice it to say as illustration here that Earth observations, be they for the management of world natural or agricultural ressources,  or for meteorological observations on one side and navigation  on another side both play an increasing role in our daily lives. Modern transportation on land, at sea or in the air all depend  on space capacities, so does our ability to predict and cope with extreme natural events like tropical storms or Earthquakes.  

Clearly not all space activities are concerned with peaceful activities meant to increase the well being of the world population. A lot of them are used to increase the dominance of some societies, be it on an economical, political, cultural or military levels. "Societies" as the word is used here means an ensemble that is large enough to develop space activities. Typically such societies are at the continental level, rather than at the national level, at least in the present Europe.

\subsection{The Present}

The dependence of our societies on space based tools means that each independent society must master these tools in order to remain independent and to foster the well being of its citizens. Failure to do so in a society implies an important dependence on the tools that other societies are willing to share. Such willingness can only be granted as long as the interests of the granting society are preserved or re-inforced.

At present the US dominate the world space activities, their space budget being about 8 times that of Europe. The military fraction of the US space budget is probably slightly over 1/2 of the total US space expenditures. It is beyond the scope of this paper to present and discuss world space budgets. The discrepancy between the European and US space budgets is, however, sufficiently large that the argument does not need further details. The large difference in expenditures  means that Europe is certainly  dependent on US  space products and technologies to a very high degree. This is painfully illustrated by the lack of a European navigation system. Several other "societies", like India or China, have taken steps to increase their space budgets in a massive way and will therefore gain independence on this aspect of their development, possibly before Europe.

Whereas space sciences have played a major role in the development of space based technologies that now form a vital element of the infrastructures of our societies, their present role is much less evident. the council of the European Space Agency (ESA) has organised a review of ESA's science programme. This lead to the "SPRT" report on the Science programme of 2007 in which one finds:
{\it
"Nowadays space science helps us to understand the evolution of the Universe and the
solar system including Earth. Space science in Europe has initially been the main driver for
the development of space technologies, which were later the basis for many applications serving
a wide range of societal needs. It provides tools and insights, which are of direct interest to mankind."
}

This means in short that whereas space science was an important element of the development of space technologie, it is not considered important anymore. Indeed policy makers, like industrial companies, require now immediate benefit from their investments in space, the days of the pioneers are long gone. Science does not offer immediate return on investment, knowledge is difficult to "count" and consequently, the space science budgets decline.
 
 \subsection{The Future}
 
 There is still an enormous volume of the parameter space in physics, astrophysics, cosmology and probably biology and elsewhere to be studied from space. To cite only one example, we are now in the very early steps in the understanding of the links between accretion onto supermassive black holes in the center of galaxies and the properties of the stellar populations that form the bulk of these galaxies. This study will require deep observations of the accretion itself, the X-ray background that represents probably the integrated accretion activity and the galaxies themselves. A study that can only be done using space instruments.
 
 The space science community in many parts of the world has the intellectual power to propose and to implement many original ideas leading to new missions and instruments. This is illustrated by the cosmic vision process in Europe and the decadal surveys in the US. The pool of knowledge and expertise available in the space science communities of the world is enormously valuable, it is, however, fragile and vulnerable and will not survive a longer decline in the space science budgets.
 
 Space science developments are in general very demanding. Observations or measurements must be optimised to the very end of the technological capacities to respond to the progressing quest for knowledge. Space sciences are therefore a unique area in which innovation is continuously pushed.
 
 Space sciences are for all these reasons a prime application to develop, train and master space tools. It should become the engine with which "societies" can gain their independence in the space sector.    Space science investments are, therefore, of prime importance to shape the relationships between "societies" in the coming decades.
 
 Space science investments offer not only a central opportunity to gain independence in all sectors of modern life, they also offer a number of side benefits. While training and mastering space application, one also learns about space and the world, one thus gains on two sides. The knowledge acquired is relatively easy to share with the public, the endeavour is therefore not confined to engineers and scientists, but is relevant for society as a whole. This knowledge also contributes to a rational approach to the world, something that is often very necessary.
 
 Space science developments are appropriate for cooperative endeavours. This means that seeking independence in space matters is not an aggressive process. This requires, however, that cooperation be made in a balanced way. Cooperation in this light is no "Ersatz" for ones own development and cannot replace large investments. Cooperation must be seen as a welcome added bonus, but may not become an essential feature of a project.

\section{Conclusion}

Astronomy has had an immense practical impact on the development of our civilisations for 1000's years and space sciences, that naturally include astronomy, have played a major part in the shaping of the modern world. Space science is, however, now not perceived as important for our development anymore. This should be massively changed and space science should become, again, a prime mover in space developments. This would allow societies to gain independence on all the areas in which space plays  a role. The effort to be done in this domain will not only be important for space applications, it will also be fruitful and and contribute to the  peaceful development of the world.

\end{document}